\documentclass[a4paper,12pt]{article}
\pdfoutput=1

\usepackage[hyphens]{url}
\usepackage{lineno,hyperref}
\usepackage{xcolor}
\hypersetup{
   colorlinks,
   linkcolor={red!50!black},
   citecolor={blue!50!black},
   urlcolor={blue!80!black}
}

\usepackage{authblk}
\usepackage[round]{natbib}

\usepackage{todonotes}

\usepackage{amsmath}
\usepackage{amssymb}
\usepackage{color}

\usepackage{multicol}
\usepackage{enumitem}

\usepackage{bm}

\usepackage[margin=0.5in]{geometry}

\begin{document}
 
\title{Effects of spatial smoothing on functional brain networks}

\author[1]{Tuomas Alak\"orkk\"o}
\author[2]{Heini Saarim\"aki}
\author[2]{Enrico Glerean}
\author[1]{Jari Saram\"aki}
\author[1,2,*]{Onerva Korhonen}

\affil[1]{Department of Computer Science, School of Science, Aalto University, Espoo, Finland}
\affil[2]{Department of Neuroscience and Biomedical Engineering, School of Science, Aalto University, Espoo, Finland}
\affil[*]{Corresponding author: Onerva Korhonen, Aalto University, Department of Computer Science, P.O. Box 15400, FI-00076 Aalto, Finland, email: onerva.korhonen@aalto.fi}

\date{}
\maketitle
 
\begin{abstract}

Graph-theoretical methods have rapidly become a standard tool in studies of the structure and function of the human brain. 
Whereas the structural connectome can be fairly straightforwardly mapped onto a complex network, there are more degrees of freedom in constructing
networks that represent functional connections between brain areas. For fMRI data, such networks are typically built by aggregating the 
BOLD signal time series of voxels into larger entities (such as Regions of Interest in some brain atlas), and determining the connection strengths between these from
some measure of time-series correlations. Although it is evident that the outcome of this procedure must be affected by how the voxel-level
time series are treated at the  preprocessing stage, there is a lack of systematic studies of the effects of preprocessing on network structure. Here, we focus on 
the effects of  \emph{spatial smoothing}, which is a standard preprocessing method for fMRI. We apply various levels of spatial smoothing
to resting-state fMRI data, and measure the changes induced in the corresponding functional networks. We show that the level of spatial
smoothing clearly affects the degrees and other centrality measures of the nodes of the functional networks; these changes are non-uniform, systematic, 
and depend on the geometry of the brain. 
The composition of the largest connected network component is also affected in a way that artificially increases the similarity
of the networks of different subjects. Our conclusion is that wherever possible, spatial smoothing should be avoided when preprocessing fMRI data
for network analysis.

\end{abstract}

\begin{multicols}{2}

\section{Introduction}

A broadly accepted consensus among the neuroscience community is that the human brain consists of interconnected, functionally specialized areas. Therefore the brain is naturally modelled 
as a complex network \citep{sporns2013a, sporns2013b, wig2011}.
In this approach, the nodes of the network represent brain areas and the links depict structural or functional connections between them. The structural features of the network may 
then help to understand the function of the
human brain. 

Application of network methods to the analysis of functional magnetic resonance imaging (fMRI) data has revealed that the brain has a non-random, hierarchical core-periphery structure. 
Network analyses have also facilitated identifying the hubs, \emph{i.e.}~the most central areas of the brain. Functional brain networks have been reported to change with age and between
health and disease as well as between different cognitive tasks. For reviews, see \citep{papo2014a, bassett2009, sporns2013b}. 

It has been criticized that in fMRI studies in general, the choice of analysis parameters is often justified insufficiently and reported incompletely \citep{carp2012secret}. Unfortunately,
this is the case for functional brain
network studies as well. It is safe to say that we do not know what effects the different data acquisition and preprocessing methods have on the structure of functional
brain networks. Indeed, factors that affect the reliability of the fMRI network studies have lately become a subject of
discussion \citep{aurich2015, shirer2015, shehzad2009, andellini2015, telesford2010, braun2012, hayasaka2013functional}. 

In the present study, we concentrate on \emph{spatial smoothing}, a commonly applied preprocessing method that may affect network properties \citep{fornito2013, stanley2013}.
In the smoothing process, the signal associated with each measurement voxel is redefined as the average of the signals 
of the voxel itself and its neighbors; typically, a smoothing kernel is applied when averaging.  Spatial smoothing has traditionally belonged to the standard set of fMRI preprocessing 
methods when the General Linear Model (GLM) is used as the analysis 
paradigm: smoothing by a Gaussian kernel ensures that data fulfills the Gaussianity assumption of the model \citep{mikl2008effects}. Spatial smoothing also increases signal-to-noise ratio
(SNR), compensates for 
inaccuracies in spatial registration, and decreases inter-subject variability \citep{hopfinger2000, bennett2010, mikl2008effects, pajula2014}. Spatial smoothing is often applied 
outside the GLM paradigm as well; in this case the justification for using it is less evident. 

In seed-based functional connectivity studies, spatial smoothing has been reported to increase connection strength between voxels, measured in terms of the correlation coefficient, and
therefore lead to detection of larger clusters of voxels connected with the seed \citep{wu2011empirical, molloy2014influence}. Meanwhile, smoothing decreased differences in connectivity between different spatial resolutions
\citep{molloy2014influence}. Scheinost \emph{et al.}~\citep{scheinost2014impact} found that image smoothness correlates with seed-based connectivity and with the degree measured in networks where nodes represent measurement
voxels. Although the different image smoothness values at different voxels are mostly caused by motion artefacts \citep{scheinost2014impact}, one may expect to see similar effects if spatial smoothing applied at 
the preprocessing stage leads to different image smoothness in different parts of the brain. In the case of Regional Homogeneity, a measure of local connectivity, spatial smoothing decreased
test-retest reliability \citep{zuo2013toward}. 

The above results indicate that spatial smoothing may affect the properties of functional brain networks. However, to the best of our knowledge, the effects
of spatial smoothing on the structure and properties of region-level functional brain networks have not been investigated in detail.

We investigate how spatial smoothing affects the structure of functional brain networks with the help of resting-state fRMI data of 13 subjects measured in-house as well 
as 28 subjects from the Autism Brain Imaging Data Exhange I (ABIDE I) initiative.
For each of these subjects, we construct the resting-state functional network using anatomically defined Regions of Interest (ROIs) as network nodes.
We show that spatial smoothing has systematic and nontrivial effects on the network structure.
At the level of individual subjects, smoothing changes which nodes appear as the most central \emph{hubs} of the network, and alters the structure of the \emph{largest connected component} that forms the core of the network.
At the group level, smoothing makes networks of different subjects more similar,
which may hinder comparisons between groups of subjects, for example patients and neurotypicals.
Based on these results, we advise against using spatial smoothing in the analysis of ROI-level functional brain networks.

\section{Methods}

\subsection{Subjects}

The in-house data used in the present study are from 13 healthy, right-handed subjects (11 females, 2 males, age 25.1 $\pm$ 3.9, mean $\pm$ std). 
They all had normal or corrected-to-normal
vision, and none of them reported a history of neurological or psychiatric disease. All subjects volunteered for the study and 
gave a written, informed consent according to the Declaration of Helsinki. Subjects were compensated for their participation. The study was approved by the Research Ethics Committee of Aalto University.

\subsection{Data acquisition}

Functional magnetic resonance imaging (fMRI) data were acquired with a 3T Siemens Magnetom Skyra scanner in the AMI Centre
(Aalto Neuroimaging, Aalto University, Espoo, Finland). A whole-brain T2*-weighted EPI sequence was collected with the 
following parameters: TR = 1.7s, 33 axial slices, TE = 24 ms, flip angle = 70 $^{\circ}$, voxel size = 3.1 x 3.1 x 4.0 mm$^3$, matrix size
64 x 64 x 33, FOV 198.4 x 198.4 mm$^2$. Data were collected from both an emotion-processing task (reported in \cite{nummenmaa2014, saarimaki2015})
and a resting-state session of approximately 6 minutes (215 time points). In the resting-state condition, subjects were instructed to lay still with their eyes open and
gaze fixated to a gray background image, and avoid falling asleep. We used only the resting-state data in the present study.

Structural MR images with isotropic 1 x 1 x 1 mm$^3$ voxel size were acquired using a T1-weighted MP-RAGE sequence.

\subsection{Preprocessing of the data}

For preprocessing of the fMRI data, we used FSL \citep{smith2004advances, woolrich2009, jenkinson2012} and an in-house
Matlab toolbox, BraMiLa (\url{https://version.aalto.fi/gitlab/BML/bramila}). First, 3 first frames of each subject's data were removed in order to minimize the error caused by scanner 
transient effect. This left time series of 212 timepoints for further analysis. Then, the preprocessing pipeline included slice timing correction, motion correction by MCFLIRT \citep{jenkinson2002},
and extraction of white matter and cerebro-spinal fluid (CSF). Functional data were co-registered to the anatomical image with FLIRT (7 degrees of freedom), registered to MNI152 template (12
degrees of freedom), and downsampled to voxels of 4 x 4 x 4 mm$^3$. Signals were linearly detrended, and signals from white matter and CSF were regressed out from the data.

Expansion of motion parameters was extracted from the data with linear regression (36 Volterra expansion based signals, \cite{power2014}) in order to control for motion artifacts.
As head motion is a possible source of artifacts in connectivity studies \citep{power2012}, framewise displacement was calculated for each subject. However, the framewise displacement
was for all subjects under the suggested threshold of 0.5 mm. Therefore, no scrubbing was performed.

In order to further eliminate artifacts, voxels that were located at the boundary of the brain and the skull and had mean signal power less than 2\% of the individual's mean signal power
were excluded from the analysis.

\subsection{Spatial smoothing}

In spatial smoothing, the time series of each voxel is redefined as an average of the time series of neighboring voxels, weighted by a smoothing kernel:

\begin{equation}
x_{i}=\frac{\sum_{j}G_{i}(j)x_{j}}{\sum_{j}G_{i}(j)}, \label{eq:smoothing}
\end{equation}

where $x_{i}$ denotes the time series of voxel $i$, $G_{i}(j)$ is the value at voxel $j$ of the smoothing kernel $G_{i}$ centered at voxel $i$, and the summation is over all voxels. For the majority of these voxels, $G_{i}(j)\approx0$.

In the present study, we used a Gaussian kernel. Spatial smoothing was always applied after 
other preprocessing steps and before  network extraction. We used three Gaussian kernels with different full width at half
maximum (FWHM): 5 mm, 8 mm, and 12 mm. As a reference, we used non-smoothed data (FWHM 0 mm).

The kernel sizes selected for the present study are commonly used among the fMRI community. Further, they correspond to what has been recommended
in the literature. Some researchers have suggested that  kernel size ``should approximate the size of the underlying signal or evoked response'' that would be approximately 3-5 mm on
the cortex \citep{hopfinger2000}. On the other hand, others think that kernel size should be 2-3 times the voxel size \citep{pajula2014, mikl2008effects}.

\subsection{Regions of Interest (ROIs)}

We divided the cortex into 96 anatomical Regions of Interest (ROIs). The ROIs were from the HarvardOxford (HO) atlas 
(\url{http://neuro.debian.net/pkgs/fsl-harvard-oxford-atlases.html}, \cite{desikan2006automated}) at 30 \% probability level (meaning that in the group used to create the parcellation,
a voxel belongs to the ROI that it is associated with in 30 \% or more of the subjects). The ROIs did not overlap, \emph{i.e.} each voxel belonged to one ROI only.

We are aware of the important role that cerebellum and subcortical areas have in the brain function (for an extensive review, see~\cite{koziol2009}). However, in the present work 
we wanted to follow the  pipelines that are commonly adopted in connectomics. These pipelines often exclude the cerebellum and subcortical areas, and therefore these areas were not
included in our analysis.

ROI time series were defined as the average over the time series of the voxels in the focal ROI:

\begin{equation}
  X_{I} = \frac{1}{N_{I}}\sum_{i\in I}x_{i}, \label{eq:roiseries}
 \end{equation} 

where $I$ is the focal ROI, $N_{I}$ is the size of the ROI $I$ (defined as the number of voxels in the ROI), and $x_{i}$ is the time series of voxel $i$.
The sizes of the ROIs used in the present study varied between 5 and 857 with the mean ROI size being 141.58$\pm$147.46 (mean$\pm$std). Median of the ROI size was 88, and 
the majority of the ROIs consisted of approximately 100 voxels. Details about the sizes of ROIs can be found in Supplementary Table.

\subsection{Network extraction}

We used the ROIs as the nodes of functional brain networks of each subject. 
We used the Pearson
correlation coefficient between the time series of each pair of ROIs to quantify the link weights. This resulted in a symmetrical adjacency matrix $A$, where the element $A_{I,J}$ 
indicated the strength of correlation between ROIs $I$ and $J$. We set the diagonal of the adjacency matrix to $A_{I,I}=0$ in order to exclude from the network self-links that
would contain no useful information. 

We used the full adjacency matrix that contains the correlations between all pairs of ROIs to investigate if spatial smoothing has different effects on links of different weight 
and physical length. To this end, we defined the physical length of a link as the Euclidean distance between the centroids of the ROIs connected by the link.

For further analysis, we thresholded the adjacency matrix to remove weak links. Low-weight links correspond to correlations that are probably too weak to be of any functional significance; further, retaining only a small number of the strongest connections provides a view on the most essential network structure. In order to obtain a network with density $d$, we removed links that were weaker than the $1-d$th weight percentile by setting the corresponding element in the adjacency matrix to 0. The weights of links that exceeded this threshold were set to 1, which yielded a binarized network.

We used the binarized networks to study how spatial smoothing affects the degrees and eigenvector centralities of nodes, since these measures have originally been defined for binary networks. 
Both  measures are commonly used to identify the most important, central nodes of the network (see section~\ref{methods:centrality}). Further, we analyzed the effects of spatial smoothing on the structure of the 
largest connected component (see section~\ref{methods:lcc}) of the binarized network.

It is not straightforward to define a single correct density for the functional brain network analysis (for discussion, see \cite{kujala2016graph}). Sometimes this problem can be overcome by investigating
a range of densities \citep{alexander2012discovery, lord2012changes}. In the present work, the behavior of
centrality measures was qualitatively the same across a range of densities from 5\% to 10\%; results reported in this article are obtained at the density of 10\% (456 links). 
For analysis of the largest connected component, we thresholded the network to a lower density of 3\% (137 links);
at this density, we observed that there is a component that is clearly larger than all others without yet spanning the entire network. 
At both densities, the network was relatively sparse
and therefore its structure was sensitive to small changes in link weights.

\subsection{Averaging and comparison of correlation coefficients}\label{methods:z-transform}

In order to investigate how spatial smoothing affected links of different weight and physical length, we needed to average correlation coefficients over subjects.
This averaging of ROI-ROI correlations was done by first Fisher Z-transforming the correlation coefficients $r$:
\begin{equation}
z = \frac{1}{2} \ln \Big( \frac{1+r}{1-r} \Big)  = \text{arctanh}(r) \label{Zt}.
\end{equation} 

The results were then averaged, and finally the values were inversely transformed back to the interval $[-1,1]$. Z-transformation decreases the bias
when averaging correlation coefficients \citep{silver1987averaging}. 

To measure the difference of two correlation coefficients, they were first Z-transformed and then subtracted.

\subsection{Centrality measures} \label{methods:centrality}

In order to study the effect of spatial smoothing on the structure of functional brain networks, we utilized two measures of node centrality: degree and eigenvector centrality. Both of 
these measures are commonly used among neuroscientists to identify the most central and important nodes, \emph{hubs}, of 
functional brain networks.

The degree of a node is defined as the number of neighbors of the node, \emph{i.e.} the number of other nodes it is directly connected with. The degree provides a simple estimate of the centrality of the node: nodes with many neighbors can be considered 
more central than those with few neighbors. Therefore, the degree can be used to define the hubs of the functional brain network \citep{rubinov2010complex}.

However, changes in the degree of a node alone do not tell how spatial smoothing has changed the centrality of the node relative to other nodes. Therefore, in order to investigate if spatial smoothing changes the "hubness" of the network nodes, we observed changes in both the degree and the degree rank of each ROI. To obtain the degree rank, we ordered ROIs in descending order by degree. Changes in the degree rank of a ROI reflect changes in its "hubness" in relation to the rest of the network.

Eigenvector centrality is a generalization of degree that also takes into account the degrees of the node's neighbors, and recursively the degrees of the neighbors of neighbors, and so on.
As stated above, a node with many neighbors can be considered as central. However, it is possible that these neighbors have low degrees, making the node less central despite its high degree. Eigenvector centrality corrects for this, and therefore measures how central the node is globally. In functional brain networks, eigenvector centrality emphasizes
the central clusters of the network \citep{lohmann2010eigenvector}.

Eigenvector centrality can be calculated iteratively from the adjacency matrix of the network. For details, see Supplementary Methods.

For investigating the centrality metrics, we wrote an in-house Python script that utilizes the NetworkX network analysis package for Python \citep{hagberg-2008-exploring}.

\subsection{Largest connected component} \label{methods:lcc}

In order to obtain a broader picture of the changes in the network structure, beyond the level of nodes, we investigated how 
spatial smoothing affects the largest connected component (LCC) of the network. In a (connected) component, every 
node can be reached from every other node by following the links of the network. In a sparse network, there can be 
several separate components; if the existence of a link is taken as indicative of functional interaction, nodes in separate components cannot influence one another. The LCC, defined as the component with the largest number of nodes, can be seen as the core of the network. 

First, we identified the LCC separately for the brain network of each subject. Then, for each node, we calculated the fraction of subjects that had that particular node in the LCC of their network.

For studying changes in the largest connected component, we used a set of in-house Python scripts (see section~\ref{methods:centrality}). 

\subsection{Network visualizations}

The HO atlas offers anatomical coordinates of the centroids of the ROIs. We used the projection of these coordinates
to the horizontal plane 
to visualize the network structure and the values of network metrics in each ROI.
We made minor adjustments to the coordinates in order to avoid overlap between ROIs while approximately retaining the anatomical position of each ROI.

\subsection{ABIDE data}

In order to ensure that our results are not explained by any particular feature of our in-house dataset and can be generalized to other datasets, we repeated all analysis of the present
article using a second, independent dataset. This dataset, to which we will now on refer as the ABIDE data, is part of the Autism Brain Imaging Data Exchange I (ABIDE I) project \cite{di2014autism}.
It contains resting-state fMRI of 28 healthy controls. For further details of the ABIDE data, see Supplementary Methods.

\section{Results}

\subsection{Spatial smoothing increases weights of short links} \label{results:short_links}

In the present work, we adopted a common approach for constructing  functional brain networks: the nodes of the network depict ROIs and link weights are defined as the Pearson correlation 
coefficients between the ROI time series. It is common to threshold weighted networks before further analysis, so that only a given percentage of the strongest links remain. 
Therefore, changes in the location of the strongest links of the full weighted network may lead to dramatical changes in the structure of the thresholded network.

\begin{figure*}[]
\centering
 \includegraphics[width=1\linewidth]{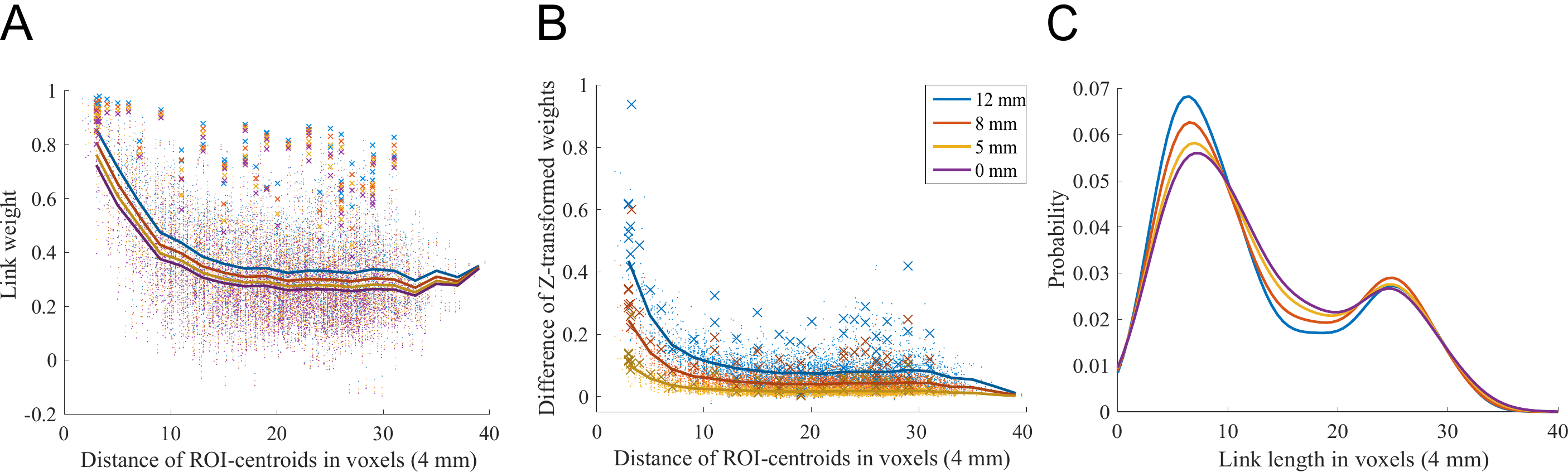} 
\caption{Spatial smoothing makes short and strong links stronger. A) ROIs that are close to each other have, on average, higher correlations, as seen from the dependence of link 
weights on the distance of the centroids of the ROIs. Smoothing increases the majority of the correlations. ROI-ROI correlations are averages over 13 subjects.
Lines are bin averages and crosses mark correlations between the same areas in the different hemispheres. Colors indicate smoothing kernel size (see panel B for legend).}
\label{main:links}
\end{figure*}

 In order to see if spatial 
smoothing has similar effects on all links, we investigated the weight of each link as a function of its physical length. The strongest links of the network are physically short 
(Fig.~\ref{main:links}A), as expected on the basis of earlier studies \citep{alexander2012anatomical, stanley2013} \footnote{The only exception are the links connecting same ROIs in opposite 
hemispheres (say, left and right frontal poles). This has been reported in the literature earlier \citep{anderson2010decreased}.}. These short links are affected by the spatial smoothing more strongly than longer links: the weights of short links increase the most (Fig.~\ref{main:links}B). Therefore, spatial smoothing alters the distributions of link lengths in thresholded networks 
(Fig.~\ref{main:links}C; $d = 10\%$).

Spatial smoothing mixes signals across ROI boundaries, between close voxels that belong to neighboring ROIs. This mixing is limited in range to the kernel width, and therefore voxels that are separated by longer distances are not directly affected. Hence, spatial smoothing increases the weights of physically short links the most.

\subsection{Spatial smoothing increases the degrees of small ROIs} \label{results:roi_size}
 
\begin{figure*}[]
\centering
 \includegraphics[scale=0.6]{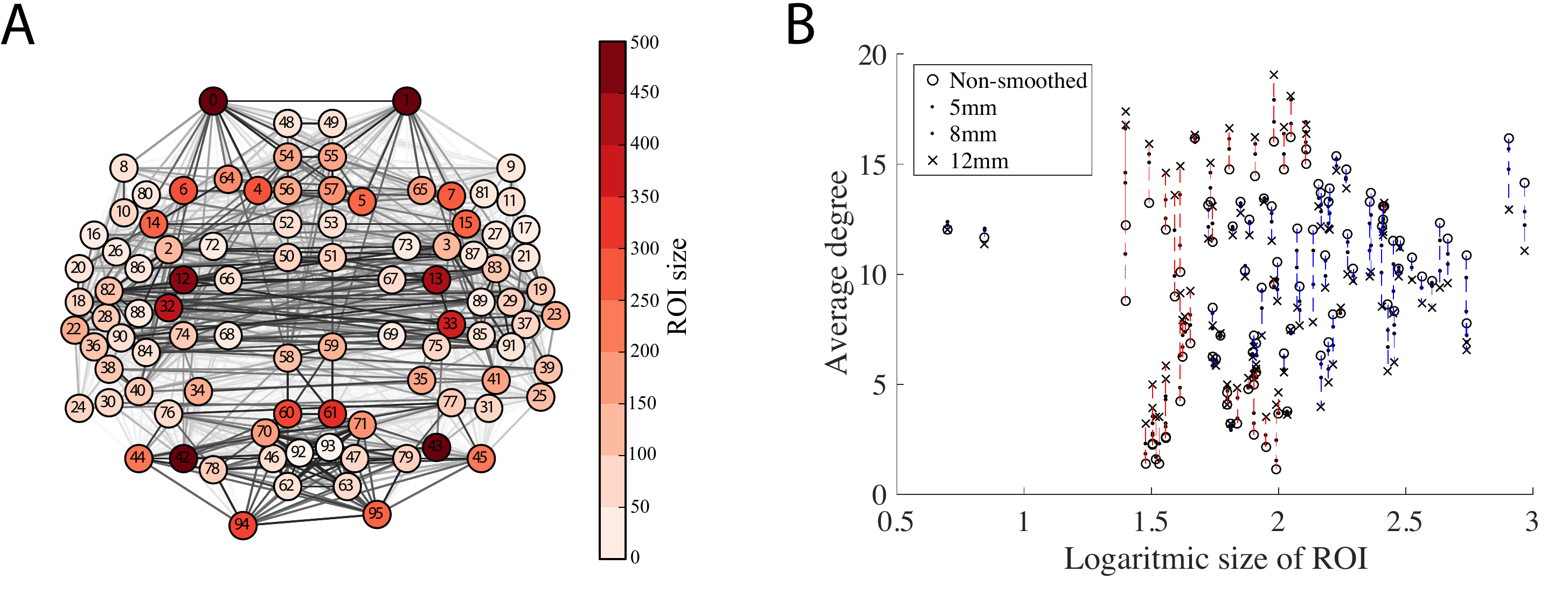}
\caption{Spatial smoothing has different effects on ROIs of different size. A) Locations and sizes of the 96 ROIs in the HarvardOxford parcellation. Locations are slightly 
shifted from anatomical locations to avoid overlap. Many small ROIs are located in temporal lobes or surrounding areas.
B) Spatial smoothing increases the degrees of small ROIs (red dashed lines) whereas the degrees of larger ROIs decrease (blue dashed lines) with the increasing FWHM of the smoothing 
kernel. ROI degrees are averaged across 13 subjects.}
\label{main:roisize}
\end{figure*}

Next, we investigated whether the effects of smoothing on nodes are uniform. Since the size of ROIs varies widely in the HarvardOxford (HO) parcellation used in the present study 
(Fig.~\ref{main:roisize}A), we ask if spatial smoothing affects ROIs of different size differently.

As stated above (section~\ref{results:short_links}), spatial smoothing mixes signals of voxels across ROI boundaries and therefore increases link weights between ROIs that are spatially 
adjacent. The signal of each ROI is a mix of the signals of voxels deep inside the ROI and the mixed voxel signals originating from the ROI boundary area (\emph{i.e.} voxels adjacent 
to at least one voxel in a different ROI). In small ROIs, this boundary area is relatively large when compared to larger ROIs. This suggests that spatial smoothing has different 
effects on the network connectivity of small and large ROIs.

In order to test this hypothesis, we investigate the degrees of ROIs, \emph{i.e.}~their numbers of network neighbors, in networks thresholded to 10$\%$ density. Here, the degree is 
determined by link weights: only links with high enough weight pass the threshold and contribute to node degree. Fig.~\ref{main:roisize}B displays the degrees as a function of ROI size 
when different levels of spatial smoothing are applied. As expected, the degrees of small ROIs are increased: signal mixing across ROI boundaries leads to higher weights of links of 
small ROIs. Since the total number of links in the thresholded network is constant, the degrees of larger ROIs decrease as the width of the smoothing kernel increases. The Pearson correlation
coefficient between ROI size and degree change quantifies this observation (FWHM5: $r=-0.55$, $p\ll 10^{-5}$; FWHM8: $r=-0.53$, $p<10^{-5}$; FWHM12: $r=-0.57$,
$p<10^{-5}$): degrees of small ROIs increase, while for large ROIs the degree change is negative.

\subsection{Network centrality measures are distorted by spatial smoothing} \label{res:network_measures}

\begin{figure*}[]
\centering
 \includegraphics[scale=0.9]{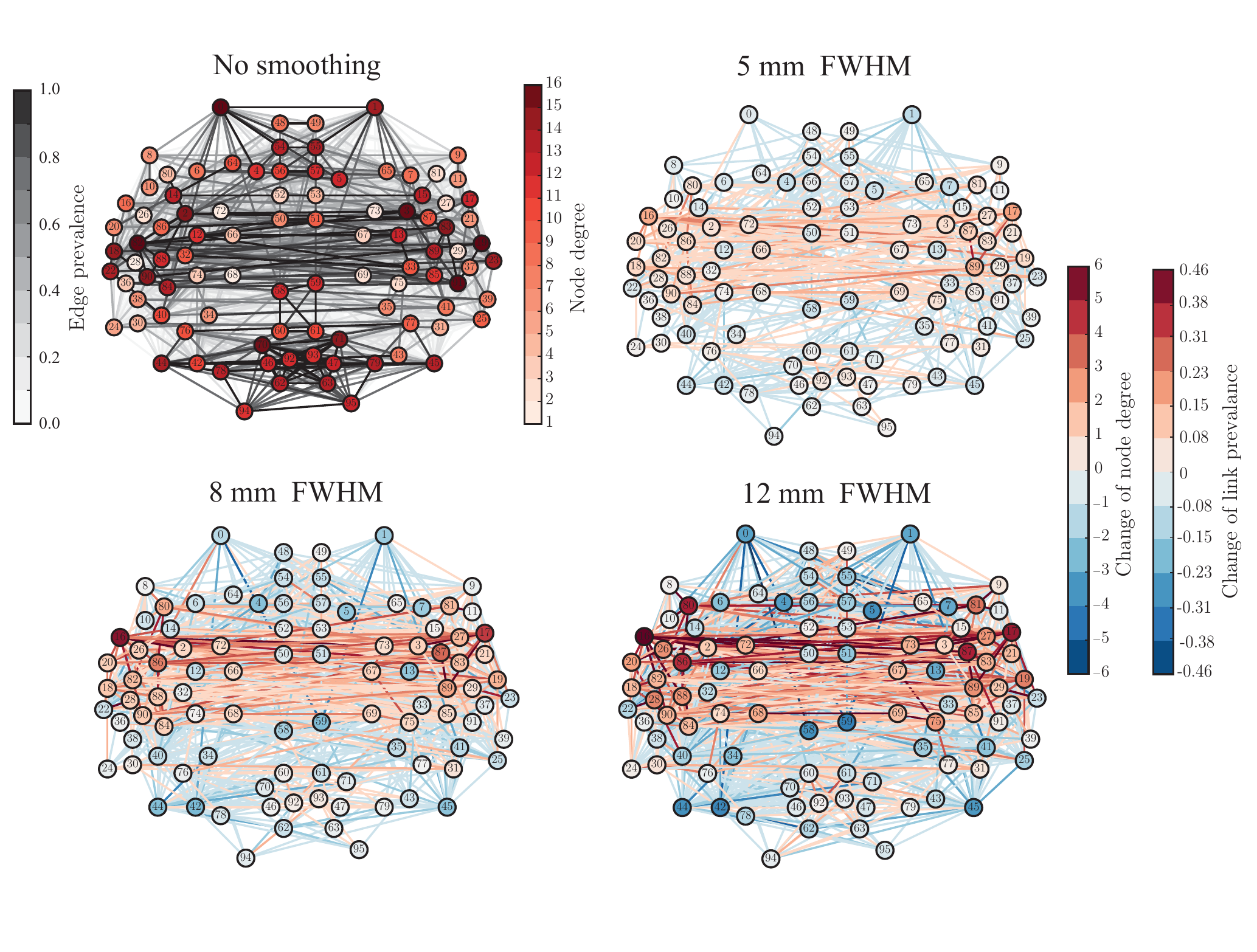}
\caption{The changes in degree due to spatial smoothing are concentrated around temporal lobes. The top left panel shows the degrees of the ROIs in the network obtained without smoothing 
applied. The other panels display differences between the network for non-smoothed data and networks resulting from smoothing the voxel-level signals using kernels with different FWHMs
(5mm, 8mm, and 12mm).
The degree values of the non-smoothed network are subtracted from the degree values of the smoothed networks; 
red (blue) color indicates increase (decrease) of degree in the smoothed network.
The colors of links indicate the change in prevalence, \emph{i.e.} fraction of subjects, out of 13, that had a given link present in their thresholded network. Networks are thresholded 
to 10\% link density. All degrees are averages over the networks of 13 subjects.}
\label{main:deg}
\end{figure*}

\begin{figure*}[]
\centering
 \includegraphics[scale=0.9]{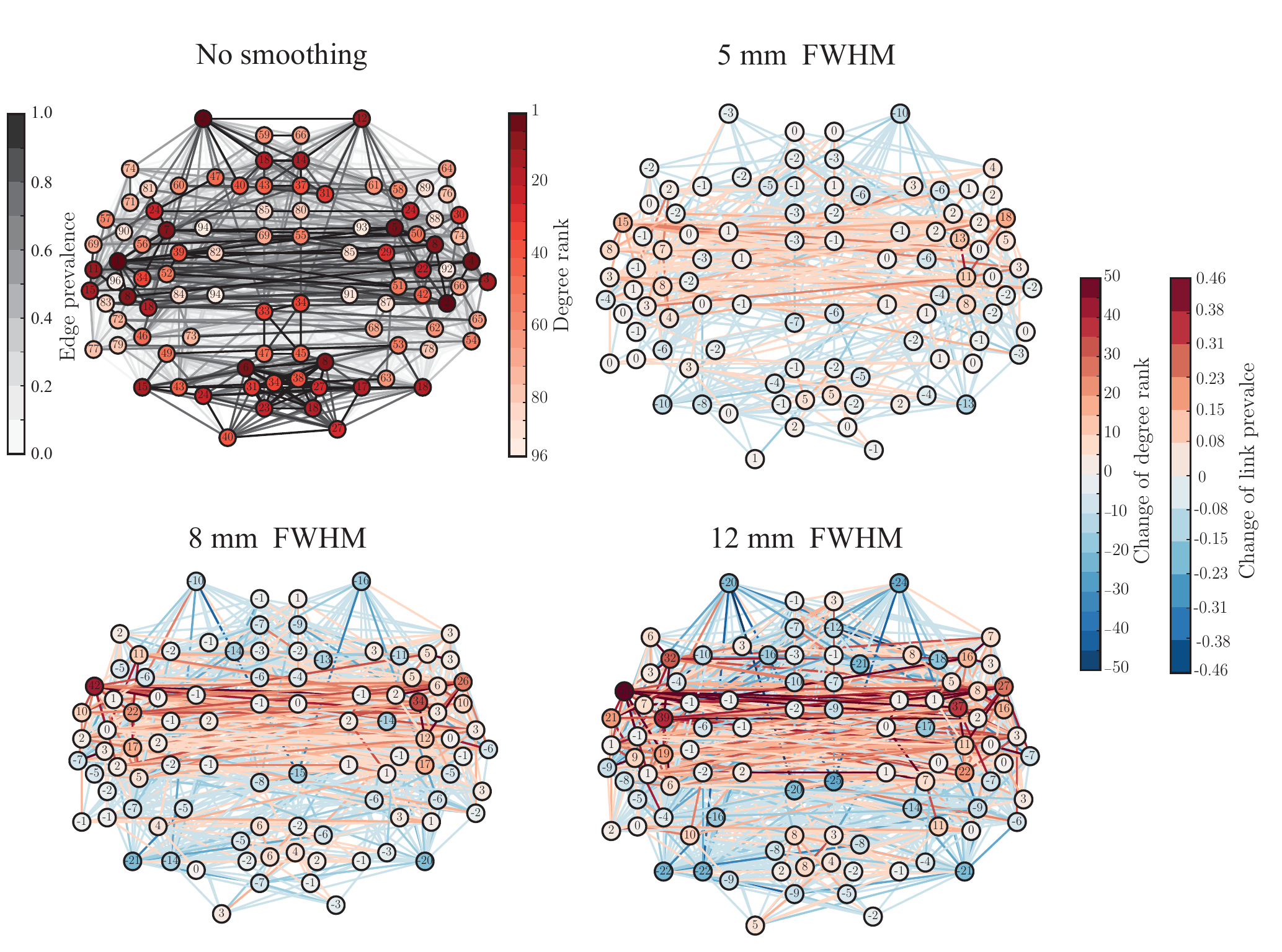}
\caption{The "hubness" of the nodes in the temporal lobes, measured in terms of their degree ranks, is increased by smoothing.
The values of ranks and rank changes are shown as node labels. The networks corresponding to smoothing kernels of 
FWHM 5mm, 8mm, and 12mm display differences as compared to the network for non-smoothed data, similarly to Fig.~\ref{main:deg}.}
\label{main:rank}
\end{figure*}

\begin{figure*}[]
\centering
 \includegraphics[scale=0.9]{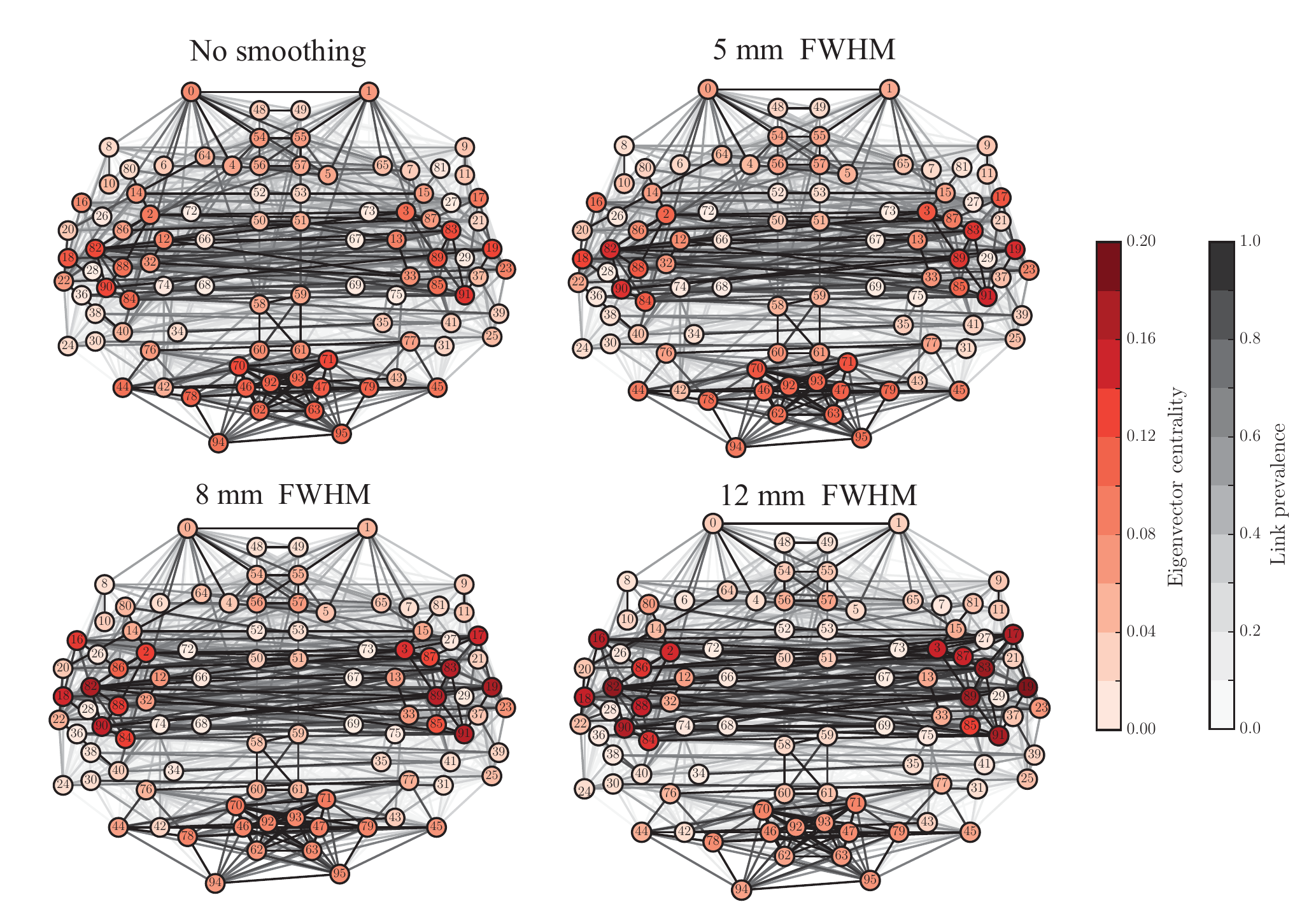}
\caption{Changes in eigenvector centralities of nodes due to smoothing reflect the changes in degree and degree ranks, as the largest increase is concentrated around the temporal lobes. 
In contrast to Figs.~\ref{main:deg} and \ref{main:rank}, node and link colors indicate absolute values of eigenvector centrality and link prevalence instead of differences.}
\label{main:eig}
\end{figure*}

Many network properties of nodes are heavily affected by node degree. Because smoothing has non-uniform effects on the degrees of ROIs, we expect to see non-uniform changes in their 
other network properties as well. We focused on the relationship between network properties and the anatomical layout of the ROI network. For each ROI, we computed the most commonly 
used centrality measures: degree as above (Fig.~\ref{main:deg}), degree rank (Fig.~\ref{main:rank}), and eigenvector centrality (Fig.~\ref{main:eig}), for different levels of smoothing. 

In a network thresholded to a fixed density (in this case, 10\%), spatial smoothing cannot change the mean degree. However, spatial smoothing changed the shape of the degree distribution:
while both the maximum and minimum degrees increased, the median degree decreased, which increased the skewness of the distribution (for details, see Supplementary Results and Supplementary Table).

This indicates that there are more ROIs with decreasing than with increasing degree, \textit{i.e.} the degree of a few
ROIs increases and the degree of most of the other areas decreases to compensate for this change. Indeed, when we visualized the values of these measures at locations corresponding to 
the anatomical coordinates of the ROI centroids (Fig.~\ref{main:deg}), the largest changes due to smoothing were observed in some ROIs in the temporal lobes and their vicinity.
In contrast, connections were weakened or lost between most other areas. 

The degree and eigenvector centrality were strongly correlated at all levels of smoothing (Pearson correlation coefficient FWHM0: $r=0.91$, $p \ll 10^{-5}4$; FWHM5: $r=0.91$, $p \ll 10^{-5}$; 
FWHM8: $r=0.90$, $p \ll 10^{-5}$; FWHM12: $r=0.90$, $p \ll 10^{-5}$). Therefore, it is not surprising that we obtained mostly similar results as for degree (Fig.~\ref{main:eig}; for details, 
see Supplementary Results and Supplementary Table).  

Spatial smoothing dramatically changes
the ``hubness'' of some nodes (Fig.~\ref{main:rank}): the degree ranks of some temporal, frontal, and parietal ROIs increased while they decreased for some occipital ROIs and midline regions
as well as for the frontal poles. 
In particular, the degree ranks increased for left and right superior temporal gyrus (anterior division), left and right middle temporal gyrus, left frontal operculum gyrus,
right parietal operculum gyrus, and left and right planum polare. Areas that decreased most in degree rank included the left and right frontal pole, right superior frontal gyrus,
superior and inferior division of left lateral occipital cortex, inferior division of right lateral occipital cortex, and anterior division of left and right cingulate gyrus.

ROIs that are hubs of the network are often assumed to be central in the transfer and processing of information during the given task. Because of this, 
the observed changes in the degree ranks may dramatically change the interpretation of the functional roles of different ROIs in the network.

\subsection{Smoothing may disconnect brain areas in the functional networks}

\begin{figure*}[]
\centering
 \includegraphics[scale=0.9]{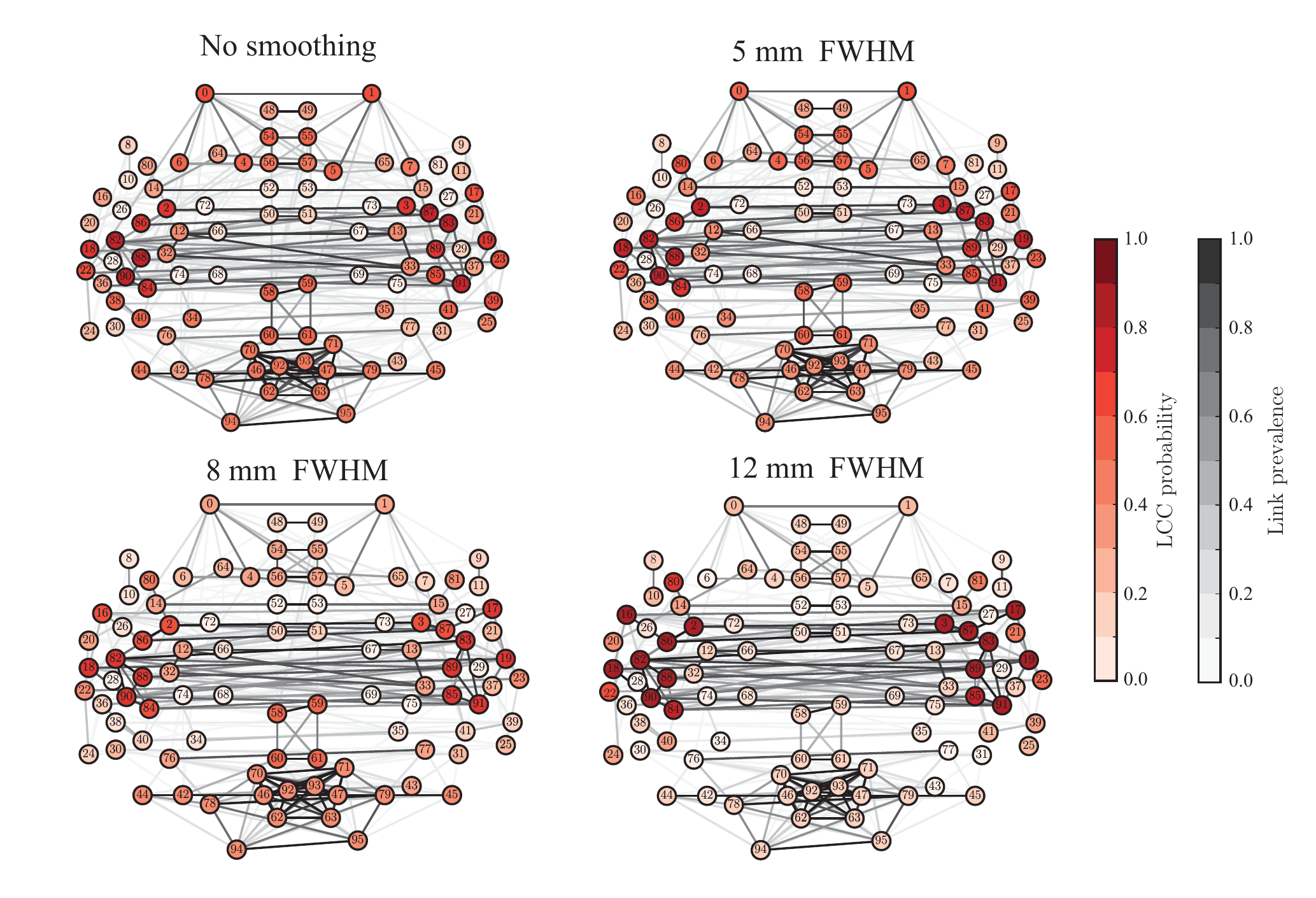}
\caption{Spatial smoothing disconnects the frontal and occipital ROIs from the LCC. Node colors indicate how frequently a certain ROI belongs to the LCC in the networks of 13 subjects.
Networks are thresholded to 3\% density.}
\label{main:lcc}
\end{figure*}

So far, we have investigated the effects of spatial smoothing at the level on single nodes and links of the functional brain networks. Next, we ask
whether spatial smoothing also changes the overall structure of the network.  
To this purpose, we extracted the largest connected components (LCCs) of the networks at low network density (3\%, 137 links) (Fig \ref{main:lcc}; see section~\ref{methods:lcc}). The LCC 
forms the functional core of the network; purely from the network point of view, all flows of information are constrained to take place within connected components only. Therefore, 
if some set of nodes drops out from the LCC, the interpretation of its functional role changes dramatically. 

In particular, we investigated the probability of nodes to belong to the LCC across subjects, \emph{i.e.} the fraction of subjects that have the ROI in
their LCC. Without spatial smoothing, a small number of ROIs had a high probability of belonging to the LCC; at the same time, many nodes had only moderate probability and were spatially spread out, indicating a high level of variation in network structure across subjects.
However, spatial smoothing decreased this variation: for a FWHM of 12 mm, the LCCs of virtually all participants comprised ROIs 
from the temporal lobes and neighboring areas. To the contrary, the occipital and frontal ROIs
that in the absence of smoothing had non-zero LCC probabilities were cut out from the LCC when smoothing was applied.

The significance of this result is that when the network is thresholded to low density, retaining only the strongest connections, 
smoothing can change which parts of brain networks are included in the largest component. With higher densities,
almost all nodes belong to the LCC, and smoothing only affects how strongly different areas are connected.

\section{Discussion}

\subsection{Spatial smoothing changes the identity of hubs of the functional brain network}

In this study, we have investigated whether the use of spatial smoothing alters the structure of functional brain networks. It has been known that spatial smoothing affects 
the properties of voxel-level networks, but these effects were suggested to be smaller at the ROI level \citep{van2008small, zalesky2012use}. However, our results show that spatial 
smoothing has significant effects on ROI-level networks built using the anatomical HarvardOxford atlas, resulting in artefacts and network features that depend on the parameters 
of the applied smoothing process. 

Applying spatial smoothing means averaging voxel signals, both within and across the boundaries of ROIs. Therefore, smoothing makes the time series of  ROIs more similar.
However, the effects of spatial smoothing are not limited to increasing the correlation between time series of physically adjacent ROIs. Rather, smoothing has also highly
nontrivial effects at the level of the whole network.

In particular, we have observed increased degree and eigenvector centrality in small ROIs. Because the degree and eigenvector centrality are commonly used to identify the hubs of the 
functional brain network \citep{rubinov2010complex,zuo2012network}, this means that the choice to use or not to use spatial smoothing during the preprocessing may produce 
different sets of nodes that are considered as hubs.
 
As smoothing increases the centralities of small ROIs, they are more likely to be considered as hubs. This is at least partially explained by their size. These ROIs have large 
boundary area relative to their volume, and therefore their time series are the ones that are most affected when voxel signals originating at different ROIs are mixed by smoothing.

\subsection{Spatial smoothing changes the structure of functional brain networks} 

As stated above, spatial smoothing increases the correlation between the time series of ROIs, which translates to increased link weights in the ROI network. However, this 
effect is not uniform across the links: the weights of physically short links grow the most, while longer links are less affected. Therefore, when the network is thresholded to a 
constant density, smoothing increases the proportion of short links.

The short links are typically located in tightly connected clusters of brain areas. To adopt the terminology of \cite{tononi1994measure} and \cite{sporns2013b}, spatial smoothing 
increases the segregation and decreases the integration of functional brain networks: connectivity within local clusters is increased whereas connections between such clusters get weaker. 
However, effective information processing in the brain requires  both integration and segregation \citep{sporns2013b}, and especially 
the longer and weaker links between local clusters are of crucial importance \citep{gallos2012small}. It is exactly these links whose strength and importance are downplayed by 
spatial smoothing.

\subsection{Spatial smoothing decreases inter-subject variation}

The largest connected component (LCC) presents the core of the functional brain network. In non-smoothed data, LCCs differ between the networks
of different subjects, whereas after smoothing the LCCs are more similar for all subjects. In other words, spatial smoothing reduces the diversity between subjects. 
This may be a desired effect in some studies. However, is there really a core of the functional brain network that is universal across subjects? After spatial smoothing, ROIs may 
belong to the LCC not because of their true centrality but because of the side-effects of smoothing. 

The decrease of inter-subject variation also
implies that spatial smoothing may decrease differences between subject groups that are supposed to be different, for example between diseased subjects and healthy controls. This
can lead to biased results in studies where \emph{e.g.} modular structures of the networks are compared between groups.

\subsection{Is spatial smoothing necessary for  ROI-level functional networks?}

Spatial smoothing is often used as a part
of the preprocessing pipeline prior to functional brain network analysis partly for historical reasons: the advantages of spatial smoothing in standard GLM analysis are beyond dispute and
smoothing therefore belongs to the standard set of fMRI preprocessing tools. Further, the effects of spatial smoothing, mainly the increase of SNR due to averaging voxel
signals, have  been thought to be advantageous in network neuroscience also. 

It is worth noting that in the ROI approach, a lot of signals are averaged in any case in order to form the time series that represent ROI activity.  Further, unlike spatial smoothing, 
this averaging does not mix signals that belong to different ROIs.
Therefore, one may question what advantage could be gained by the additional averaging due to spatial smoothing.

\subsection{Conclusion}

ROI-level resting-state functional brain networks are affected by spatial smoothing. 
Spatial smoothing has complex effects on the structure and properties
of the networks, including possible over-emphasis of strong, short-range links, changes in the identities of hubs of the network,
and decreased inter-subject variation. The ROI approach already includes averaging, independent of spatial smoothing.
Therefore, there is no specific reason for applying spatial smoothing.

\section*{Acknowledgements}

We acknowledge the computational resources provided by the Aalto Science-IT project. We thank Marita Kattelus and Athanasios Gotsopoulos for their help in data acquisition.

\section*{Author contributions}

TA, OK, and JS designed the study. HS supplied data. TA, EG, and OK contributed tools for preprocessing and analysis. TA performed the analysis. TA, OK, and JS wrote the manuscript with
the help of comments of all authors.

\end{multicols}

\section*{Abbreviations}

\begin{description}[leftmargin=*]
 \item[ABIDE] Autism Brain Imaging Data Exhange
 \item[BOLD] blood-oxygen-level dependent
 \item[CSF] cerebro-spinal fluid
 \item[EPI] echo planar imaging
 \item[fMRI] functional magnetic resonance imaging 
 \item[FWHM] full width at half maximum
 \item[FOV] field of view
 \item[GLM] general linear model
 \item[HO] HarvardOxford
 \item[LCC] largest connected component
 \item[MR] magnetic resonance
 \item[ROI] region of interest
 \item[SNR] signal-to-noise ratio
 \item[TE] echo time
 \item[TR] repetition time
\end{description}

\bibliographystyle{apalike}
\bibliography{references}

\pagebreak

\begin{center}
 \textbf{\large Effects of spatial smoothing on functional brain networks}
\end{center}

\setcounter{equation}{0}
\setcounter{figure}{0}
\setcounter{page}{1}
\setcounter{section}{0}
\makeatletter
\renewcommand{\theequation}{S\arabic{equation}}
\renewcommand{\thefigure}{S\arabic{figure}}
\renewcommand{\bibnumfmt}[1]{[S#1]}
\renewcommand{\citenumfont}[1]{S#1}

\section{Supplementary Methods}

\subsection{Iterative algorithm for calculating eigenvector centrality}

We calculate the eigenvector centrality with an iterative algorithm utilizing the adjacency matrix of the network.
First, eigenvector
centrality of each node of the network is initialized to 1. Then, at each timepoint $t$ the eigenvector centrality of the
focal node $i$ is defined as

 \begin{equation}
  c^{\mathrm{eig}}_{i}(t)=\sum_{j=1}^{N}a_{ij}c^{\mathrm{eig}}_{j}(t-1), \label{eq:eig-centrality}
 \end{equation}

where $N$ is the total number of nodes in the network, $a_{ij}$ is the element of the adjacency matrix $\bm{A}$ that corresponds
to the link between nodes $i$ and $j$, and $c^{\mathrm{eig}}_{j}(t-1)$ denotes the eigenvector centrality
of node $j$ at timepoint $t-1$. We can write this in the matrix form:

 \begin{equation}
  \bm{c^{\mathrm{eig}}}(t)=\bm{Ac^{\mathrm{eig}}}(t-1)=\bm{A}^{t}\bm{c^{\mathrm{eig}}}(0). \label{eq:eig-centrality_matrix}
 \end{equation}

Next, we can express this in terms of eigenvectors $\bm{v_{i}}$ and eigenvalues $\lambda_{i}$ of $\bm{A}$:

\begin{equation}
 \bm{c^{\mathrm{eig}}}(t)=\lambda_{1}^{t}\sum_{i=1}^{N}b_{i}\left[\frac{\lambda_{i}}{\lambda_{1}}\right]^{t}\bm{v_{i}},
 \label{eq:eig-centrality_lambda}
\end{equation}

where $\lambda_{1}$ is the largest eigenvalue of $\bm{A}$ and $b_{i}$s are a set of multipliers. Now, when time increases to infinity,

\begin{equation}
 \lim_{t\to\infty}\bm{c^{\mathrm{eig}}}(t)=\lambda_{1}^{t}b_{1}\bm{v_{1}}. \label{eig-centrality_final}
\end{equation}

So, after normalization, the eigenvector centrality vector is the eigenvector that corresponds to the
largest eigenvalue of the adjacency matrix.

\subsection{ABIDE data}

In order to ensure that the results obtained in the present article are not caused by any feature unique to our in-house dataset, we repeated all the analysis for a second,
independent dataset. This dataset has been published as a part of the Autism Brain Imaging Data Exchange I (ABIDE I) initiative \citep{di2014autism}. From now on, we will refer
to this dataset as the ABIDE data.

\subsubsection*{Subjects}

The abide dataset contains the data of 28 healthy control subjects. These subjects were measured as a part of the ABIDE I initiative; 19 subjects at California Institute of Technology (Caltech) and
9 subjects at Carnegie Mellon University (CMU). In the present study, subjects from both measurement sites were pooled in order to form a single dataset.

ABIDE subject IDs of the 19 subjects measured at Caltech were 51475, 51476, 51477, 51478, 51479, 51480, 51481,
51482, 51483, 51484, 51485, 51486, 51487, 51488, 51489, 51490, 51491, 51492, and 51493. Out of these subjects, 15 were male and 4 female. Their age ranged between 17 and 56.2 years.
15 of the subjects were right-handed and one left-handed, while the handedness of 3 subjects was ambiguous. The subjects reported no history of either autism spectrum disorders (ASD) or
other psychiatric or neurological disease nor family history of ASD. The Caltech data has been described in detail in \citet{tyszka2014}.

ABIDE subject IDs of the 9 subjects measured at CMU were 50657, 50659, 50660, 50663, 50664, 50665, 60666, 60667, and 50668. All the CMU subjects were male. Their age varied between
21 and 40 years. 8 of them were right-handed and one had ambiguous handedness. None of these subjects had reported history of ASD or other psychiatric or neurological disease.

\subsubsection*{Data acquisition}

The MRI and fMRI data of the subjects measured at Caltech were acquired with a 3 Tesla Magnetom Trio device (Siemens Medical Solutions, NJ, USA). Structural MR images were acquired
with a T1-weighted MP-RAGE sequence with $1\times1\times1 \text{mm}^3$ isotropic voxel size. The resting-state fMRI data were measured as a whole-head T2*-weighted EPI 
sequence with the following parameters: TR = 2.0 s, TE = 30 ms, flip angle = 75 $^{\circ}$, voxel size = $3.5\times3.5\times3.5 \text{mm}^3$, matrix size = 64 x 64 x 34, 
FOV = 256 x 256 mm$^2$. The length of the time series was 3 minutes (150 time points). During the measurement, the subjects were instructed to lie still and keep their eyes closed but
prevent falling asleep.

For subjects measured at CMU, MRI and fMRI data were acquired with a 3 Tesla Magnetom Verio device (Siemens Medical Solutions, NJ, USA). Structural MR images were acquired with a 
T1-weighted MP-RAGE sequence
with isotropic voxel size of $1\times1\times1 \text{mm}^3$. The resting-state fMRI data were measured with a whole-head T2*-weighted EPI sequence with the following parameters:
TR = 2.0 s, TE = 30 ms, flip angle = 73 $^{\circ}$, voxel size = $3.0\times3.0\times3.0 \text{mm}^3$, matrix size = 64 x 64 x 20, FOV = 192 x 192 mm$^2$. The length of the time series
was 10.4 min (320 time points). During the scanning, subjects lay still with their eyes closed in a room with lights shut off.

\subsubsection*{Preprocessing and analysis}

The preprocessing and analysis pipeline of the ABIDE dataset was identical to the one applied on our in-house dataset. For details, see the Methods section of the main article. The measurement parameters
slightly differed between the Caltech and CMU subjects. Therefore, in order to avoid any unpredictable effects caused by differences in data acquisition, the anatomical MR images of 
all subjects were registered to MNI152 standard template and resampled to voxel size of $4\times4\times4\text{mm}^3$ prior to creating the group-level masks for obtaining ROIs. Further,
time series of different subjects -- and possibly of different lengths -- were never averaged during the analysis. Instead, functional brain networks were extracted separately for each
subject, and all averaging over subjects was done only after
extracting the networks.

Note that the number of ROIs differend between our in-house dataset and the ABIDE data: the ABIDE data contained only 92 anatomical ROIs. This is because the locations of 
ROIs 27, 28, 73, and 74 (left and right inferior temporal gyrus, anterior division, and left and right temporal fusiform cortex, anterior division) did not overlap across all ABIDE 
subjects. In the ABIDE data, the ROI size varied between 5 and 765 with mean size 127.07$\pm$124.18 (mean$\pm$std). The median ROI size was 86.5.

\section{Supplementary Results}

\subsection{Changes in centrality distributions}

In the main article (section \ref{res:network_measures}), we noticed that spatial smoothing alters the shape of the degree distribution in our in-house dataset. In a network
thresholded to a fixed density, spatial smoothing cannot change the mean degree. Instead, 
the median and standard deviation of degree decreased with the 
increasing level of smoothing (FWHM0: $k=10.19$, $\text{std}=4.22$; FWHM5: $k=10.04$, $\text{std}=4.19$; FWHM8: $k=9.62$, $\text{std}=4.17$;
FWHM12: $k=9.19$, $\text{std}=4.12$).  Meanwhile, both the minimum
and maximum degrees increased (FWHM0: $k_{min}=1.15$, $k_{max}=16.15$; FWHM5: $k_{min}=1.54$, $k_{max}=16.92$; FWHM8: $k_{min}=2.31$, $k_{max}=17.92$; FWHM12: $k_{min}=3.23$,
$k_{max}=19.08$),
and the number of degree values smaller than the mean increased. In summary, spatial smoothing increased the skewness of the degree distribution.

As one may expect, degree and eigenvector centrality were strongly correlated in our in-house dataset, and also changes in the distributions of degree and eigenvector centrality
due smoothing were mostly the same. The median eigenvector centrality decreased with the increasing level of smoothing (FWHM0: $c^{\mathrm{eig}}=0.059$, FWHM5: $c^{\mathrm{eig}}=0.054$,
FWHM8: $c^{\mathrm{eig}}=0.048$, FWHM12: $c^{\mathrm{eig}}=0.038$), and both the maximum and minimum eigenvector centrality increased (FWHM0: $c^{\mathrm{eig}}_{max}=0.077$,
$c^{\mathrm{eig}}_{min}=0.0030$; FWHM5: $c^{\mathrm{eig}}_{max}=0.15$, $c^{\mathrm{eig}}_{min}=0.0030$; FWHM12: $c^{\mathrm{eig}}_{max}=0.17$, $ c^{\mathrm{eig}}_{min}=0.0040$;
FWHM12: $c^{\mathrm{eig}}_{max}=0.19$, $c^{\mathrm{eig}}_{min}=0.0050$). However, there were also some differences. First, as the
mean of eigenvector centrality is not fixed by the network density, it decreased with increasing level of smoothing (FWHM0: $c^{\mathrm{eig}}=0.061$, FWHM5: $c^{\mathrm{eig}}=0.061$,
FWHM8: $c^{\mathrm{eig}}=0.058$, FWHM12: $c^{\mathrm{eig}}=0.056$). Further, contrary to degree, the standard deviation of 
eigenvector centrality increased with smoothing (FWHM0: $\text{std}=0.038$, FWHM5: $\text{std}=0.041$, FWHM8: $\text{std}=0.046$, FWHM12: $\text{std}=0.052$).

\subsection{ABIDE data}

\begin{figure*}[]
\centering
 \includegraphics[width=0.9\linewidth]{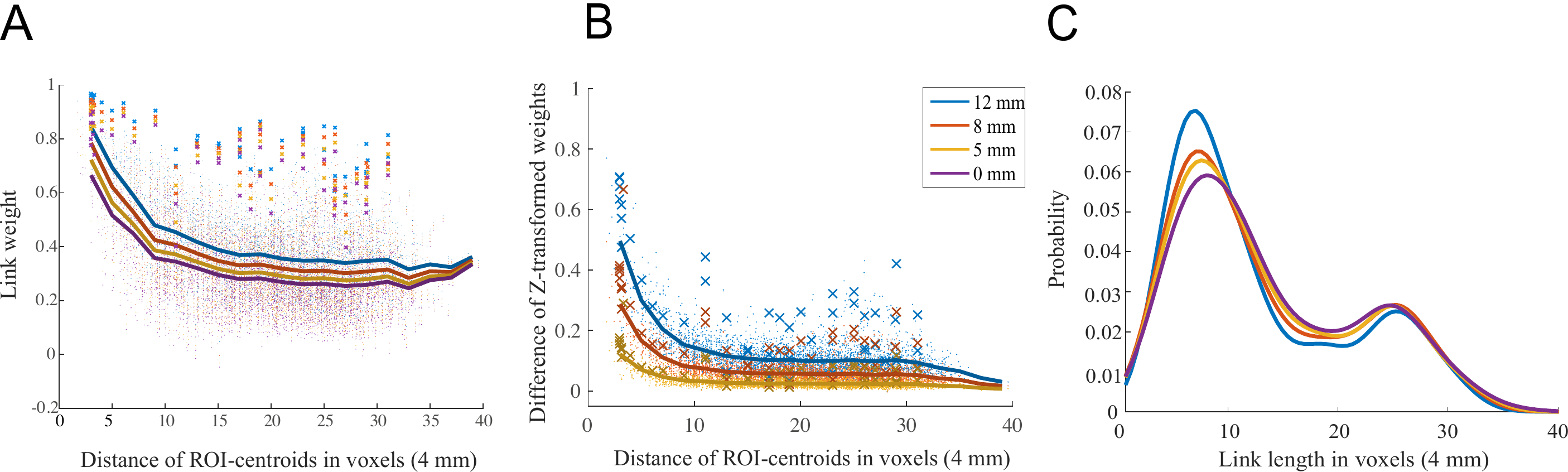} 
\caption{Short, strong links get stronger when spatial smoothing is applied. A) The strongest links connect ROIs that are physically close to each other. This dependency of the link
weight
and the physical length of a link is independent of spatial smoothing, although smoothing increases correlations between all ROIs. ROI-ROI correlations are averages over 28 participants.
Lines show bin averages, and crosses mark correlations between the corresponding anatomical areas in the different hemispheres. Colors indicate smoothing kernel size (see panel B for legend). 
B) The increase in weight due to spatial smoothing is
largest for short links. The change of link weight between each smoothing level and the non-smoothed case is calculated using Fisher's Z-transform (for details, see section~\ref{methods:z-transform} in the main
article). C) Spatial smoothing alters the distribution of the physical link length. Distributions have been calculated from the pooled data of 28 subjects in networks thresholded to
$d=10\%$.}
\label{sup:links}
\end{figure*}

We started our analysis of the ABIDE data by investigating how spatial smoothing affects the links of the functional brain network, where the nodes depict Regions of Interest (ROIs)
and links are defined as Pearson correlation coefficients between the time series of ROIs. As in the case of the in-house dataset, also in the networks extracted using the ABIDE data
the strongest links were physically short (Fig.~\ref{sup:links}A). Spatial smoothing increased the weight of these short links more than the strength of longer links (Fig.~\ref{sup:links}B). This
lead to an altered distribution of link lengths in thresholded networks (Fig.~\ref{sup:links}C; $d=10\%$).

\begin{figure*}[]
\centering
 \includegraphics[scale=0.6]{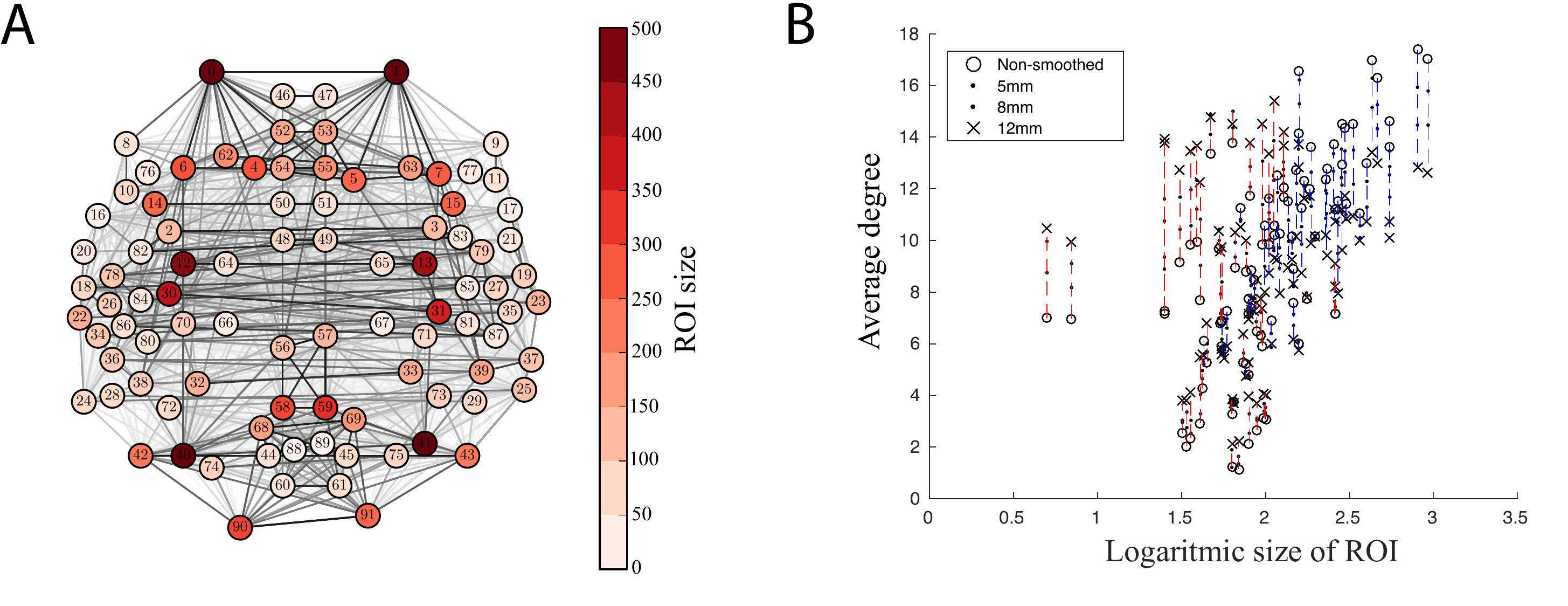}
\caption{Spatial smoothing increases degrees of small ROIs. A) Locations and sizes of the 92 anatomical ROIs from the HO parcellation. Locations are slightly 
shifted from anatomical locations to avoid overlap. Many small ROIs are located in temporal lobes or surrounding areas. B) Spatial smoothing 
increases the degrees of small ROIs (red dashed lines) whereas the degrees of larger ROIs decrease (blue dashed lines). ROI degrees are averaged over 28 subjects.}
\label{sup:roisize}
\end{figure*}

We continued by investigating if spatial smoothing changes the degree of the nodes of the functional brain networks. In the analysis of the in-house dataset, we have noticed that degrees
of small ROIs, \textit{i.e.} ROIs that contain few voxels, increased when smoothing is applied. This is true also for the ABIDE dataset: small ROIs gained new links
in networks thresholded to $d=10\%$ (Fig.~\ref{sup:roisize}). Meanwhile, the degree of larger ROIs decreased, since the number of links in the thresholded network must stay constant.
The Pearson correlation coefficient between ROI size and degree change quantifies this observation (FWHM5: $r=-0.63$, $p<10~{-5}$; FWHM8: $r=-0.64$, $p<10^{-5}$;
FWHM12: $r=-0.65$, $p<10^{-5}$): degree of small ROIs increased, while degree of larger ROIs decreased.

\begin{figure*}[]
\centering
 \includegraphics[scale=0.9]{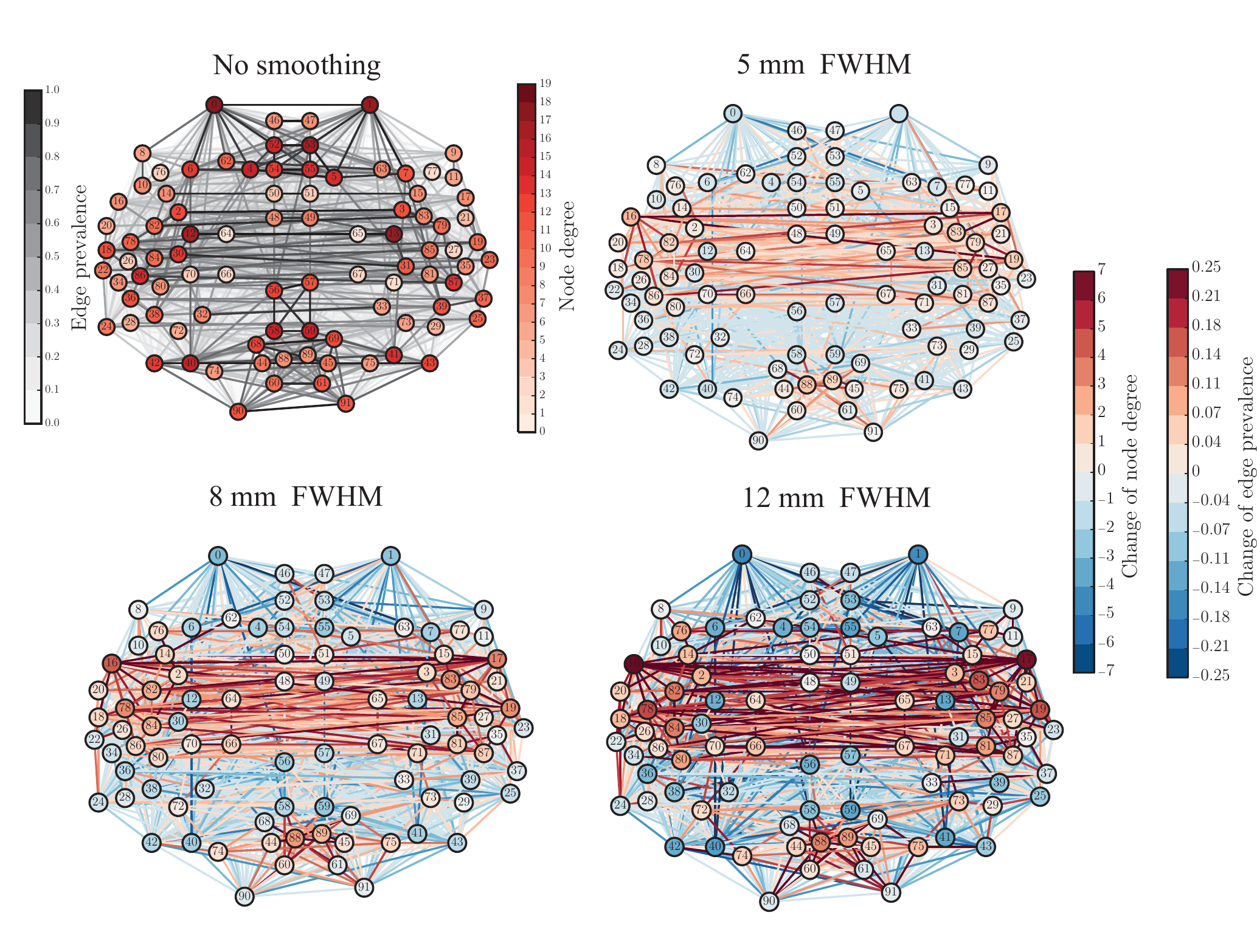}
\caption{Spatial smoothing increases degrees of temporal and occipital ROIs. The top left panel shows, as a reference, the degrees of the ROIs in the network extracted from non-smoothed
data. Rest of the panels display degree differences between the reference network and networks extracted from data that have been smoothed with different-sized
kernels (FWHM of 5 mm, 8 mm, and 12 mm). Red (blue) colors indicate increase (decrease) of degree in networks extracted from smoothed data. 
The colors of links indicate the change in prevalence, \emph{i.e.} fraction of subjects, out of 28, that had a given link present in their thresholded network. 
Networks are thresholded to 10\% link density. All degrees are averages over the networks of 28 subjects.}
\label{sup:deg}
\end{figure*}

\begin{figure*}[]
\centering
 \includegraphics[scale=0.9]{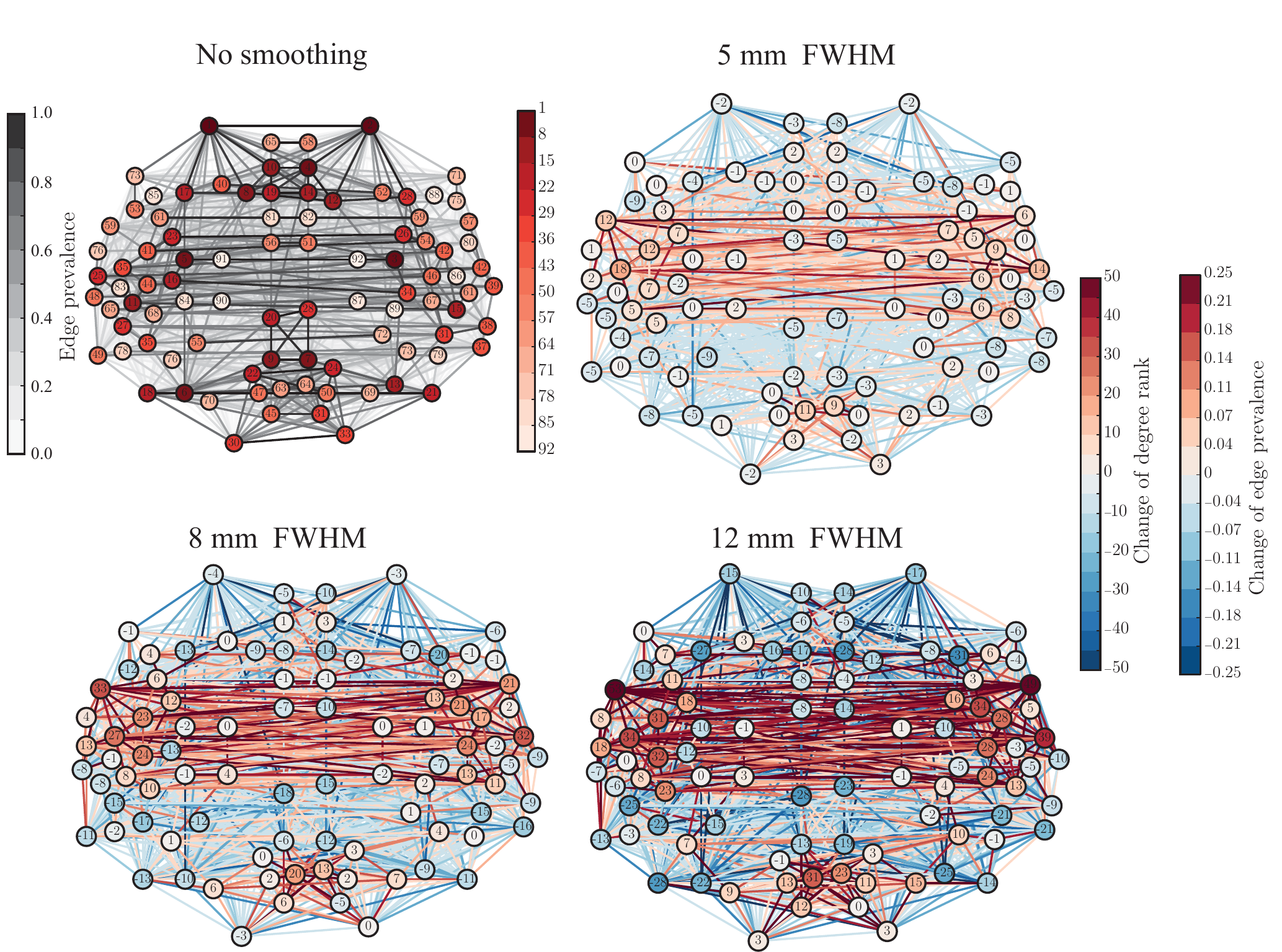}
\caption{Spatial smoothing increases the ``hubness'', measured by degree rank, of ROIs in temporal and occipital lobes. 
The values of ranks and rank changes are shown as node labels. The networks corresponding to smoothing kernels of 
FWHM 5mm, 8mm, and 12mm display differences as compared to the network for non-smoothed data, similarly to Fig.~\ref{sup:deg}.}
\label{sup:rank}
\end{figure*}

Node degree is commonly used to define the most central nodes, \textit{hubs}, of the network. Therefore, the non-uniform changes in node degree due to spatial smoothing hint that 
smoothing possibly affects the hubs of the network. To investigate this in more detail, we calculated three commonly-used centrality measures: node degree (Fig.~\ref{sup:deg}), degree
rank (Fig.~\ref{sup:rank}), and eigenvector centrality (Fig.~\ref{sup:eigenvector_centrality}) for each ROI. For detailed results, see Supplementary Table.

Similarly as in the in-house data, we obtained more negative than positive degree changes. The minimum degree increased and standard deviation decreased as a function of the width of the
smoothing kernel
(FWHM0: $k_{min}=1.14$, FWHM5: $k_{min}=1.21$, FWHM8: $k_{min}=1.64$, FWHM12: $k_{min}=2.14$).
However, contrary to the in-house data, the median degree increased 
(FWHM0: $k=9.07$, FWHM5: $k=9.54$, FWHM8: $k=9.73$, FWHM12: $k=9.36$)
and maximum degree decreased with the 
increasing level of smoothing (FWHM0: $k_{max}=17.39$, FWHM5: $k_{max}=16.28$, FWHM8: $k_{max}=15.28$, FWHM12: $k_{max}=15.39$. This indicates that the changes caused by spatial 
smoothing are nontrivial and may depend also on the properties of the dataset analyzed.

Degree and eigenvector centrality were strongly correlated also in the ABIDE dataset (Pearson correlation coefficient FWHM0: $r=0.97$, $p \ll 10^{-5}$; FWHM5: $r=0.96$, $p \ll 10^{-5}$;
FWHM8: $r=0.96$, $p \ll 10^{-5}$; FWHM12: $r=0.94$, $p \ll 10^{-5}$).
Mean and median of the eigenvector centrality decreased with the increasing level of smoothing similarly as in the in-house data
(FWHM0: mean: $c^{\mathrm{eig}}=0.066$, median: $c^{\mathrm{eig}}=0.067$; FWHM5: mean: $c^{\mathrm{eig}}=0.065$, median: $c^{\mathrm{eig}}=0.068$; FWHM8: mean: $c^{\mathrm{eig}}=0.064$,
median: $c^{\mathrm{eig}}=0.067$; FWHM12: mean: $c^{\mathrm{eig}}=0.061$, median: $c^{\mathrm{eig}}=0.059$). In standard deviation or minimum and maximum values of 
eigenvector centrality we found no clear, systematic changes (see Supplementary Table). Also the changes in the eigenvector centrality of single nodes were smaller than in the in-house data.

Similarly as in the in-house data, increase in degree rank was concentrated in temporal lobes and their vicinity. The areas, for which the degree rank increased most, included left and right
superior temporal gyrus (anterior division), right superior temporal gyrus (posterior division), left and right central operculum cortex, left and right planum temporale, left and right
Heschl's gyrus, and left supracalcarine cortex. Among the areas that decreased the most in degree rank were left and right middle frontal gyrus, left supramarginal gyrus (posterior division),
right occipital gyrus (superior division), left lateral occipital cortex (inferior division), right cingulate gyrus (anterior division), and left and right cingulate gyrus (posterior 
division).

\begin{figure*}[]
 \centering
 \includegraphics[scale=0.9]{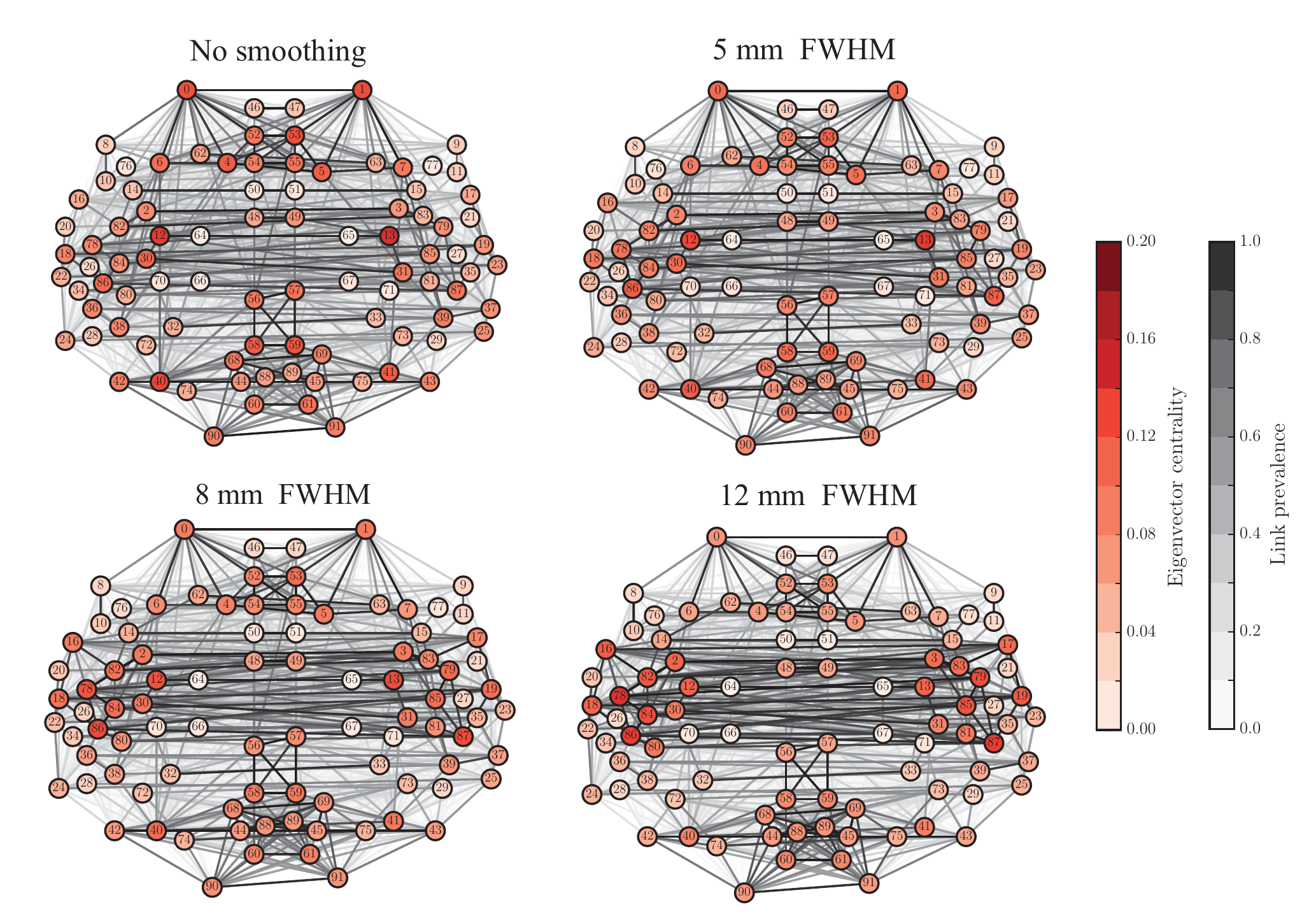}
 \caption{Spatial smoothing changes eigenvector centrality values of nodes. Similarly as in the case of degree, the centrality of ROIs located in the temporal lobes increases the most,
 while eigenvector centrality of most other ROIs decreases. In contrast to Figs.~\ref{sup:deg} and \ref{sup:rank}, node and link colors indicate absolute values of eigenvector centrality and 
 link prevalence instead of differences, for networks constructed from non-smoothed and smoothed data. Eigenvector centrality values are averages over 28 subjects.
 }
 \label{sup:eigenvector_centrality}
\end{figure*}

\begin{figure*}[]
 \centering
 \includegraphics[scale=0.9]{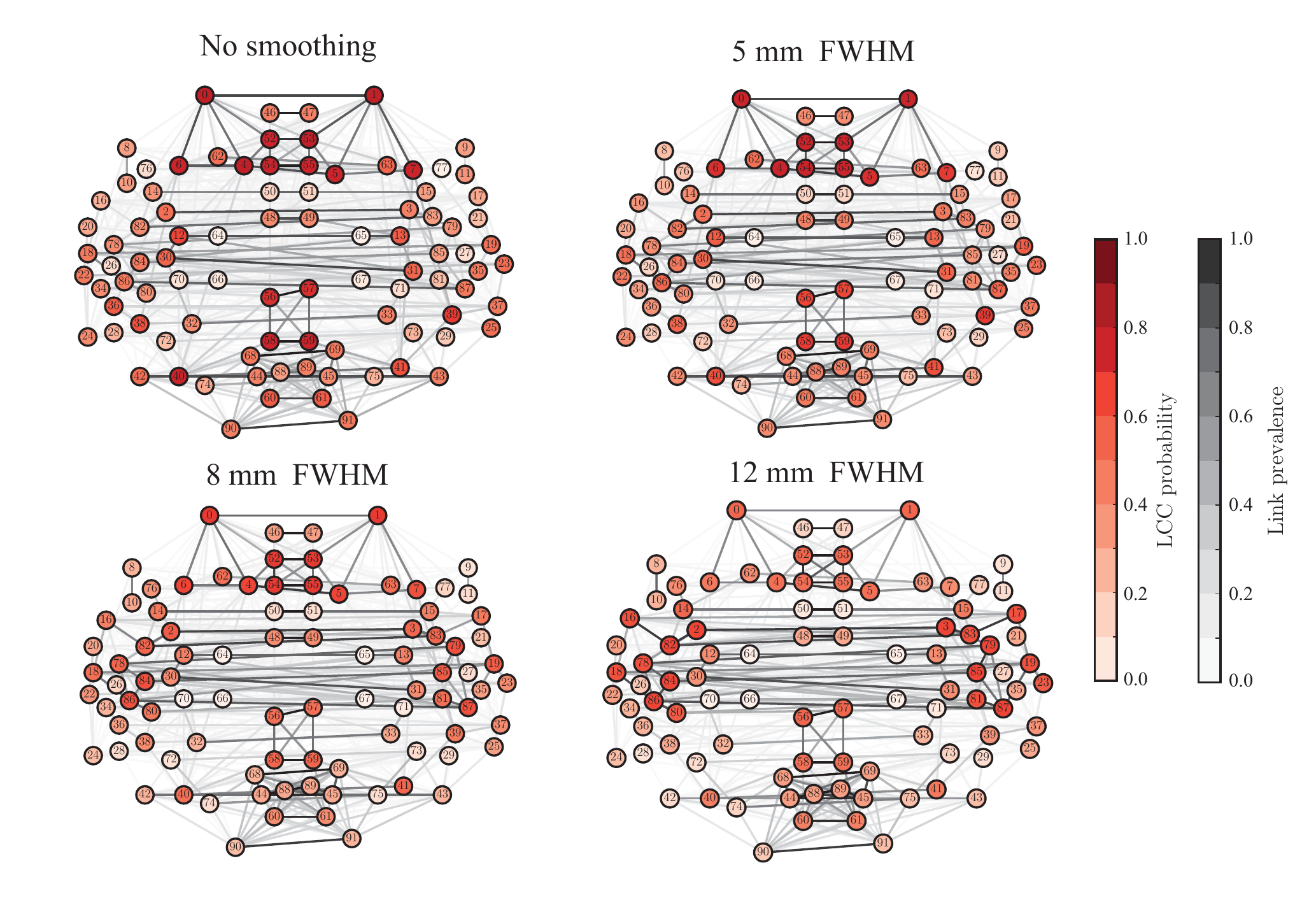}
 \caption{Spatial smoothing changes the structure of the LCC. Although the change is not as clear as in the in-house data, also in the ABIDE data occipital and frontal areas are less
 probable to belong to the LCC when smoothing has been applied.
 Node colors indicate how frequently a certain ROI belongs to the LCC in the networks of 28 subjects.
 Networks are thresholded to 3\% density.}
 \label{sup:lcc}
\end{figure*}

Finally, we asked if spatial smoothing changes the overall structure of the functional brain networks. In particular, we investigated the largest connected component (LCC) of the 
network. In the case of the in-house data, we noticed that spatial smoothing decreased the inter-subject variation in the structure of the LCC. Further, smoothing also significantly decreased the probability
of some areas, especially occipital ROIs and frontal poles, to belong to the LCC. In the case of the ABIDE data, this effect is not as visible as in the in-house data (Fig.~\ref{sup:lcc}),
although some decrease in probability of belonging to the LCC can be observed especially in the occipital lobe. One possible explanation for the differences in the effect size between
the in-house data and the ABIDE data is the number of subjects (13 vs 28). Since the ABIDE dataset contains more subjects, the original inter-subject variance may be larger than in the
in-house dataset and therefore better survive also after spatial smoothing.

\section{Supplementary Table}

The Supplementary Table is available at \url{https://github.com/onerva-korhonen/effects-of-spatial-smoothing} in Excel format (.xlsx).
In the Supplementary Table, we present detailed numerical results about the effects of spatial smoothing, separately for each Region of Interest. The table shows names and sizes of the
96 HO ROIs used to analyze the in-house dataset and the 92 ROIs used to analyze the ABIDE dataset, as well as degrees, degree ranks, eigenvector centralities, and probabilities to
belong to the LCC calculated with different levels of spatial smoothing applied.

\end{document}